\documentstyle[multicol,aps,prl,epsfig]{revtex}

\begin{document}
\title{Universality Class of Thermally Diluted Ising Systems at Criticality}
\author{Manuel I. Marqu\'es and Julio A. Gonzalo}
\address{Depto. de F\'isica de Materiales C-IV\\
Universidad At\'onoma de Madrid\\
Cantoblanco 28049 Madrid Spain\\
email: julio.gonzalo@uam.es}
\author{Jorge \'I\~niguez}
\address{Depto. de F\'isica Aplicada II\\
Universidad del Pais Vasco\\
Apdo 644 48080 Bilbao Spain}
\date{25-1-2000}
\maketitle

\begin{abstract}
The universality class of thermally diluted Ising systems, in which the
realization of the disposition of magnetic atoms and vacancies is taken from the local distribution of 
spins in the pure original Ising model at criticality, is
investigated by finite size scaling techniques using the Monte Carlo method.
We find that the critical temperature, the critical exponents and therefore
the universality class of these thermally diluted Ising systems depart markedly
from the ones of short range correlated disordered systems. Our results agree fairly
well with theoretical predictions previously made by Weinrib and Halperin
for systems with long range correlated disorder.
\end{abstract}



\vskip1pc

\begin{multicols}{2}
\narrowtext
\parskip=0cm


{\bf I. Introduction} \vskip1pcDuring last decades the systems with quenched
randomness have been intensively studied \cite{Domb}. The Harris criterion 
\cite{Harris} predicts that weak dilution does not change the character of
the critical behavior near second order phase transitions for systems of
dimension $d$ with specific heat exponent lower than zero in the pure case (the
so called P systems), $\alpha _{pure}<0\Longrightarrow \nu _{pure}>2/d$, being $\nu$ the correlation length critical exponent.
This criterion has been supported by renormalization group (RG)  
\cite{Lubensky,LubenskyII,Grinstein}, and scaling analyses \cite{Aharony}%
. The effect of strong dilution was studied by Chayes et al. \cite{Chayes}.
For $\alpha _{pure}>0$ (the so called R systems), the Ising 3D case for example,
the system fixed point flows from a pure (undiluted) fixed point towards a
new stable fixed point \cite
{Lubensky,LubenskyII,Grinstein,Aharony,Chayes,Kinzel} at which $\alpha
_{random}<0$. Recently Ballesteros et al. have used the Monte Carlo approach
to study the diluted Ising systems in two\cite{BallesterosI}, three \cite
{BallesterosII} and four dimensions \cite{BallesterosIII}. The existence of
a new universality class for the random diluted Ising system (RDIS),
different from that of the pure Ising model and independent of the average
density of occupied spin states $(p)$, is proved, using an infinite volume
extrapolation technique\cite{BallesterosII} based upon the leading
correction to scaling. The critical exponents obtained this way could
be compared with the experimental critical exponents for a random disposition of vacancies
in diluted magnetic systems \cite{Folk}.

In all cases previously mentioned frozen disorder was always produced in a
random way, that is, vacancies were distributed throughout the lattice {\bf %
randomly}. Real systems, however, can be realized with other kinds of
disorder, where the vacancy locations are correlated. In particular, long
range correlation (LRC) has been found in X-ray and Neutron Critical
Scattering experiments in systems undergoing magnetic and structural phase
transitions\cite{Shirane,ShiraneII}. This effect has been 
modeled by assuming a spatial distribution of critical temperatures obeying
a power law $g(x)\sim x^{-a}$ for large separations $x$ \cite{Altarelli}.
In general these systems behave in a way very different from
RDIS, since systems with randomly distributed impurities may be considered
as the limit case of short range correlated (SRC) distributions of the
vacancies. The basic approach to the critical phenomena of LRC systems was
established by Weinrib and Halperin \cite{Weinrib} almost two decades ago.
They found that the Harris criterion can be extended for these cases,
showing that for $a<d$, the disorder is irrelevant if $a\nu _{pure}-2>0$,
and that in the case of relevant disorder a new universality class (and a
new fixed point) with correlation length exponent $\nu =2/a$, and a specific
heat exponent $\alpha =2(a-d)/a$ appears. In contrast, if $a>d$, the usual
Harris criterion for SRC systems is recovered. LRC disorder has been studied
also by the Monte Carlo approach \cite{BallesterosIV}. In this case a
correlation function $g(x)=x^{-a}$ with $a=2$ (defects consisting in
randomly oriented lines of magnetic vacancies inside a three dimensional
Ising system) confirmed the theoretical predictions of Weinrib and Halperin.

In the present paper we will study Ising three dimensional systems where the
long range correlated dilution has been introduced as a thermal
order-disorder distribution of vacancies in equilibrium, governed by a characteristic
ordering temperature ($\theta$), in a way similar to the thermal disordering in a
binary alloy of magnetic (spins) and non-magnetic atoms (vacancies). [See
previous work \cite{Marques} and also the application
of this kind of disorder in percolation problems \cite{Stanley}]. Making
more explicit the analogy with order-disorder in alloys we may distinguish
clearly between:

i) {\bf thermally diluted} Ising system realizations (TDIS), in which the
quenched randomness is produced by considering a ferromagnetic Ising system
at $(\theta =T_{c}^{3D})$: after thermalization, the spins of the dominant
type (concentration $c\geq 0.5$) are taken as the locations of the magnetic
atoms, and the rest are taken as the magnetic vacancies. The structure of
the realization is fixed thereafter for all temperatures at which the
magnetic interactions are subsequently investigated.

ii) {\bf random} {\bf diluted} Ising system realizations (RDIS) (equivalent to $\theta >>T_{c}^{3D}$), 
also fixed thereafter for all temperatures at
which the spin interactions are investigated. Of course, we chose for
comparison a vacancy probability $p=0.5$, resulting in $c\approx 0.5.$

iii) as a third possibility, not considered here, one could investigate what
might be called an {\bf antiferromagnetically diluted} Ising system (with $\theta
<<T_{N}^{3D}$, which would lead to a disposition of
non-magnetic atoms (vacancies) strictly alternating with spins, with $c=0.5$.

So, if this ordering temperature $\theta $ determining the particular
realization is high enough, the equilibrium thermal disorder will be very similar
to the random (short range correlated) disorder of the usual previous
investigations. On the other hand, if $\theta $ happens to coincide with
the characteristic magnetic critical temperature ($T_{c}^{3D}=4.511617$) of
the undiluted system, we will have vacancies in randomly located {\bf points}, but
with a long range correlated distribution. [Note that the situation differs
markedly from that of previously studied LRC systems, in which {\bf lines} or
{\bf planes} of vacancies were considered]. The correlation of our TDIS is given
by a value $a=2-\eta _{pure},$ where $\eta _{pure}$ is the correlation
function exponent for the pure system.\ Since $d=3$ and $\eta _{pure}=0.03$
for the three dimensional Ising system, we have a long range correlated
disorder with $a=1.97<3=d$. So we are in the case where LRC disorder is
relevant and we should detect a change of universality class with respect to
the SRC case (following Weinrib and Halperin we expect for the thermally
diluted Ising system $\nu \approx 1$ and $\alpha \approx -1)$. Details
about the construction of these thermally diluted Ising systems (TDIS) can
be found in Ref.\cite{Marques}. In the present work we study the critical
behavior and the university class of three dimensional TDIS at criticality
using the Monte Carlo approach. We will compare our results with the
critical behavior of the RDIS.

The structure of the paper is as follows: In Section II we study the
dependence of the critical temperature (and of the self-averaging at
criticality) with the size of the system for both TDIS and RDIS. Once the
critical point is determined, we investigate whether or not TDIS and
RDIS belong to the same universality class. In order to proceed, we study
the critical behavior of both kinds of systems by applying finite size
scaling techniques (Section III) and by using the effective-exponent
approach (Section IV). A summary of the main results and concluding remarks
are given in Section V.

\vskip1pc{\bf II. Transition temperature and self-averaging of thermally
diluted systems} \vskip1pcFor a hypercubic sample of linear dimension $L$ and
number of sites $N=L^{d}$, any observable singular property $X$ has different
values for the different random realizations of the disorder, corresponding
to the same dilution probability $p$ (grand-canonical constraint). This means
that X behaves as a stochastic variable with average $\overline{X}$ (in the
following, the bar indicates average over subsequent realizations of the
dilution and the brackets indicate MC average). The variance would then be $(\Delta X)^{2}$,
and the normalized square width, correspondingly: 
\begin{equation}
R_{X}=(\Delta X)^{2}/\overline{X}^{2}
\end{equation}
A system is said to exhibit self-averaging (SA) if $R_{X}\rightarrow 0$ as $%
L\rightarrow \infty $. If the system is away from criticality, $L>>\xi $
(being $\xi $ the correlation length). The central limit theorem indicates
that strong SA must be expected in this case. However, the behavior of a ferromagnet at
criticality (with $\xi >>L)$ is not so obvious. This point has been studied
recently for short range correlated quenched disorder. Aharony and Harris
(AH), using a renormalization group analysis in $d=4-\varepsilon $ dimensions,
proved the expectation of a rigorous absence of self-averaging
in critically random ferromagnets \cite{AharonyII}. More recently, Monte
Carlo simulations were used to investigate the self-averaging in
critically disordered magnetic systems \cite
{BallesterosII,WisemanII,AharonyIII}. The absence of self-averaging was
confirmed. The normalized square width $R_{X}$ is an universal quantity
affected just by correction to scaling terms. LRC diluted systems are
expected to have different critical exponents and different normalized
square widths with respect to those of the usual randomly disordered systems
studied previously.

We perform Monte Carlo calculations of the magnetization and the
susceptibility $(\chi =(\left\langle M^{2}\right\rangle -\left\langle
M\right\rangle ^{2})/T)$ per spin at different temperatures for different
realizations of TDIS, and for randomly diluted systems with $p=0.5$ (restricting
to $c>0.5$) using in both cases the Wolff \cite{Wolff} single
cluster algorithm \cite{Wang} with periodic boundary conditions, on lattices
of different sizes $L=10,20,40,60,80,100$. Results for susceptibility vs.
temperature are shown in Fig. 1.
\begin{figure}
\epsfxsize=\columnwidth\epsfbox{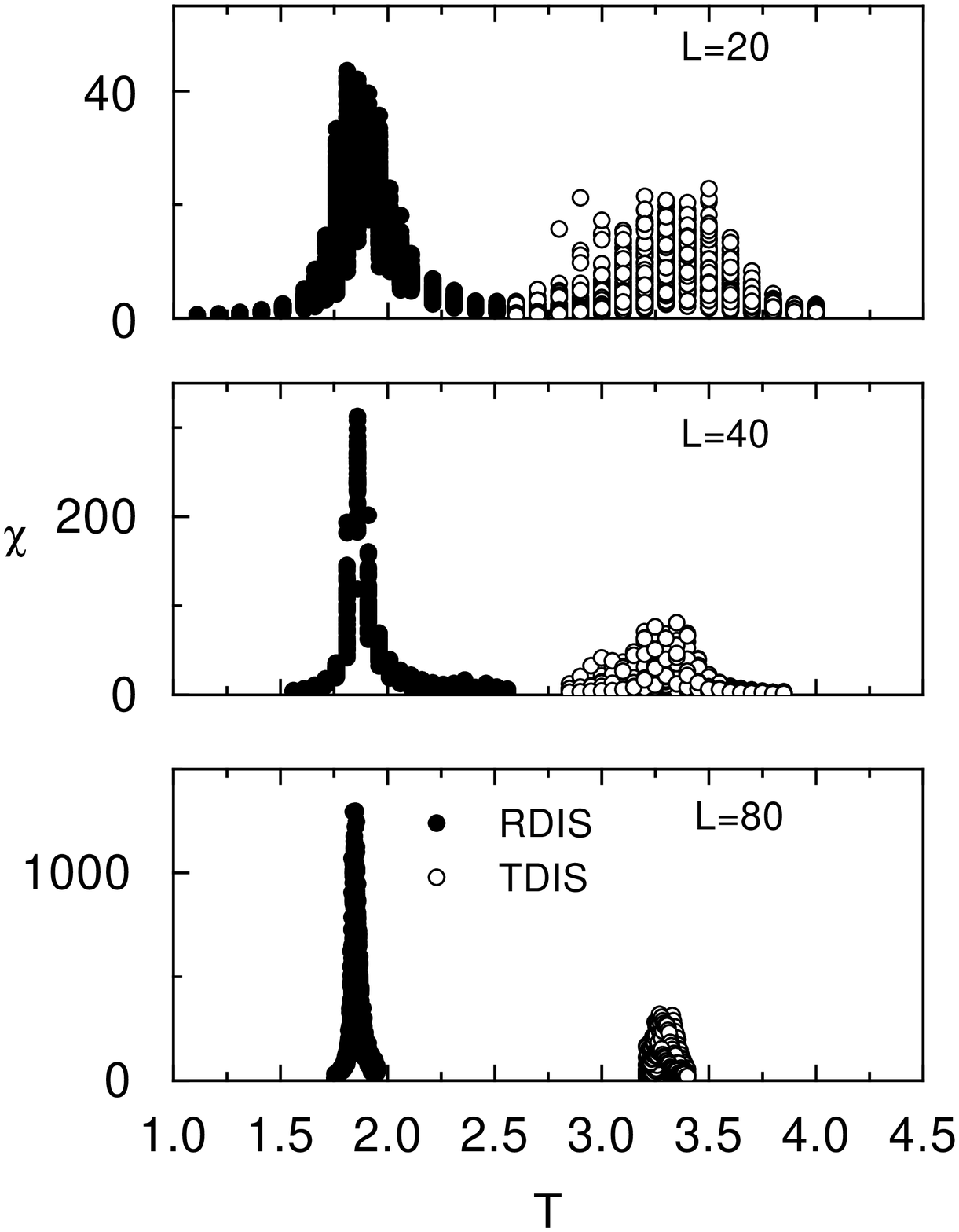}
\caption{Susceptibility $\protect\chi$ vs. temperature $T$ for random
dilution realizations ($p=0.5$) (black) and critical thermal dilution
realizations (white). The size of the systems under consideration is given
by $L=20,40,80 $.}
\label{fig1}
\end{figure}

Note how due to the existence of randomness
the susceptibility points do not collapse into a single curve, since each
realization has a different value of the critical temperature and of the
concentration. This is even more clear for small values of L and for the
critically thermal case (white points). Fig. 1 indicates that the critical
temperature of the TDIS clearly differs from that of the RDIS, and also than
the effect of the dilution on the lack of self-averaging seems to be
stronger in the thermally diluted Ising case. The critical temperature may
be obtained at the point where the normalized square width for the
susceptibility, $R_{\chi }$, reaches its maximum. A different value of the
critical temperature does not imply, of course, a different universality
class. However from the dependence of the critical temperature with the
length of the system we expect to detect a change in universality between
TDIS and RDIS following the scaling relation: 
\begin{equation}
T_{c}(L)=T_{c}(\infty )+AL^{-1/\nu }
\end{equation}
being $\nu $ the critical exponent associated with the specific system's
correlation length. This critical exponent has been determined by means of
Monte Carlo data for the random case by Ballesteros et al. \cite
{BallesterosII}. They found a value $\nu _{random}=0.683$. On the other hand
the result by Weinrib and Halperin \cite{Weinrib} indicates that the
critical exponent expected for the thermal case should be $\nu
_{thermal}=2/a=1.015$. Fig. 2 represents the dependence of the critical
temperature with respect to the length of the systems for random and thermal
dilutions.
\begin{figure}
\epsfxsize=\columnwidth\epsfbox{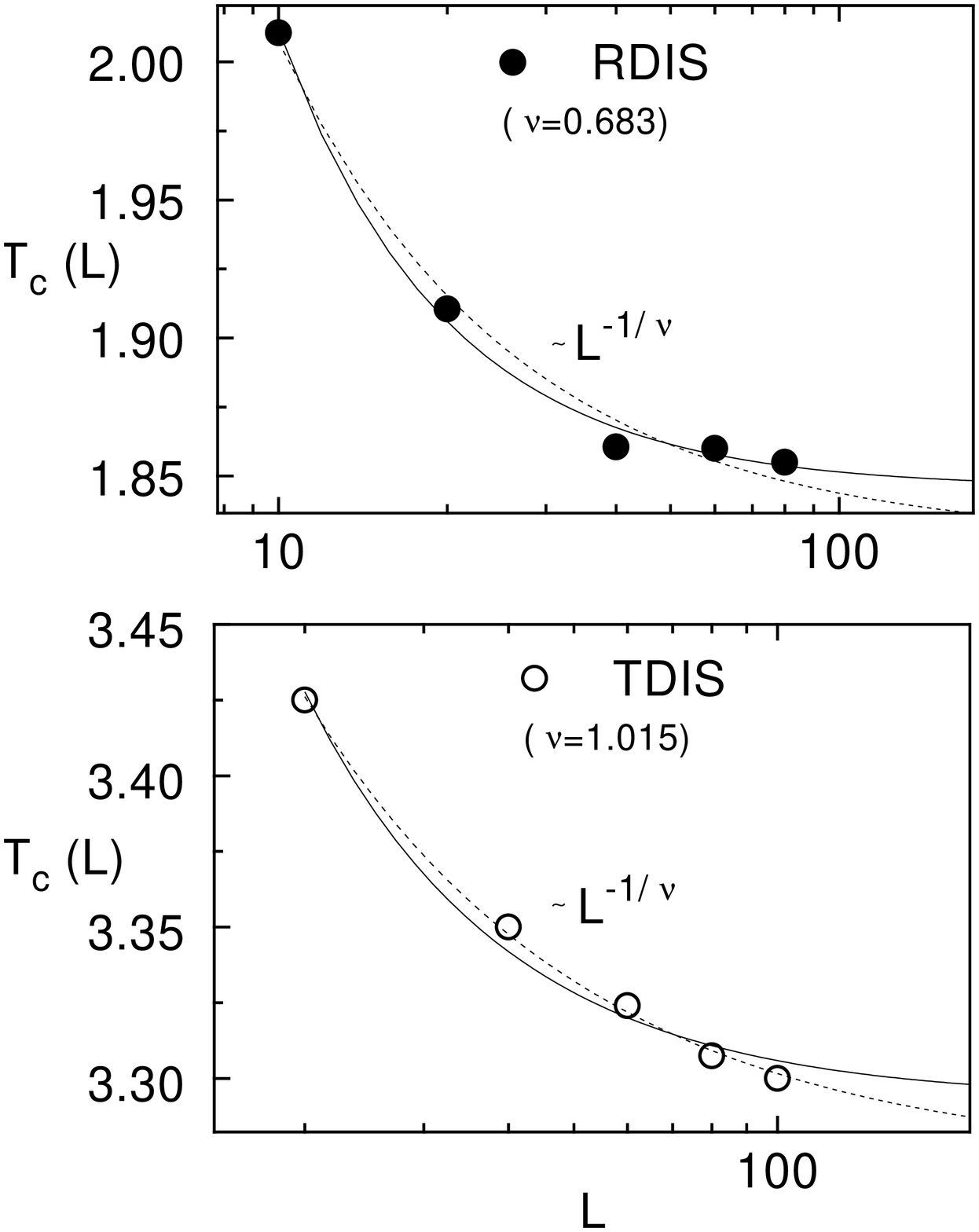}
\caption{Semi log representation of critical temperature $T_c$ vs. lenght $L$
for random dilution ($p=0.5$) (black) and for thermal dilution (white). The
continuous line is the fit obtained using $\protect\nu=0.6837 $ (short range
correlated random exponent) and dashed line the fit obtained using $\protect%
\nu=1.015$ (long range correlated exponent, a=1.97) in either case.}
\label{fig2}
\end{figure}

In both cases a fit to Eq. 2 has been performed for both
values $\nu =\nu _{random}=0.683$ (continuous line) and $\nu =\nu
_{thermal}=1.015$ (dashed line). Note how the thermal data fit better Eq. 2
for $T_{c}(L)$ using $\nu _{thermal}$, indicating a possible change in
universality class with respect to the random case. [If we fit the data
leaving all parameters free, we find $\nu _{random}\approx. 0.7$
 and $\nu _{thermal}\approx 1.2$, which are very near the
expected results]. The extrapolated values of critical temperatures for
infinite systems obtained this way are $T_{c}^{random}(\infty )=1.845\pm
0.003$ (close to the values previously obtained by Ballesteros et al.\cite
{BallesterosII}) and $T_{c}^{thermal}(\infty )=3.269\pm 0.002$ (clearly
different from $T_{c}(\infty )$ for the SRC case). Incidentally $\nu
_{thermal}$ can be compared with $\nu $ for the observed sharp component in
neutron scattering line shapes, which is around 1.3 for Tb \cite{ShiraneII}.
This point deserves more careful analysis and will be taken up elsewhere.

Once the critical temperatures are known we can perform simulations for the
magnetization and the susceptibility at criticality for several realizations
of thermal and random diluted systems in an effort to determine the value of 
$R_{\chi }$, the normalized square width for the susceptibility. We went up
to 500 realizations for $L=10,20,40$ and up to 200 realizations for $L=60,80$. Results
are shown in Fig.3.
\begin{figure}
\epsfxsize=\columnwidth\epsfysize=7.5cm\epsfbox{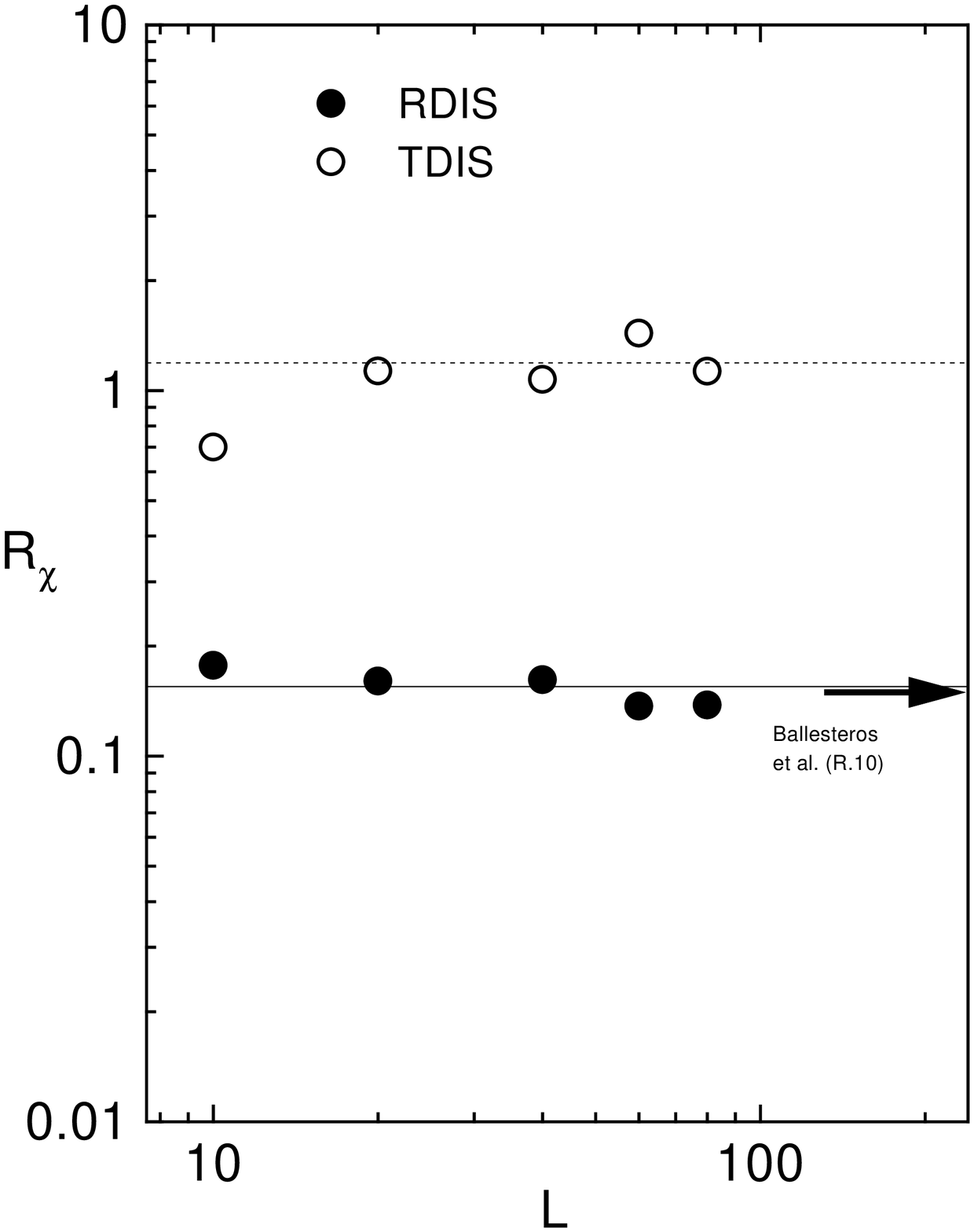}
\caption{Log-log plot of the normalized square width for the susceptibility
at criticallity $R_\protect\protect\chi $ vs. lenght $L$ for random dilution
($p=0.5$) (black) and thermal dilution (white).}
\label{fig3}
\end{figure}

The arrow represents the (concentration
independent) $R_{\chi }$ value obtained by Ballesteros et al \cite
{BallesterosII}. The straight continuous line represents the average value
obtained for random dilution data, and the straight dashed line gives the
average value obtained for thermal dilution data. Note that the TDIS
presents a lack of self-averaging around one order of magnitude larger than
the RDIS. We have already presented a similar analysis for both kinds of
dilution \cite{Marques}, but only at the critical temperature {\bf characteristic
of the random system}. Our results are not precise enough to specify
accurately the evolution of the normalized square width as a function of L
governed by corrections to scaling terms. However the average we obtain for 
$R_{\chi }^{random}=0.155$ is close to the value previously reported 
\cite{BallesterosII} by means of infinite volume extrapolations. For
the thermal case we obtain $R_{\chi }^{thermal}=1.19$, about one order of
magnitude larger than for random dilution. In this case an evolution of $%
R_{\chi }^{thermal}$ vs. $L$ given by correction to scaling terms may be also
expected, but according to Weinrib and Halperin \cite{Weinrib} the analysis would be even more complicated, due to the fact
that the long range correlated disordered systems present complex
oscillating corrections to scaling.
\vskip1pc{\bf III. Critical behavior and exponents of thermally diluted
systems} \vskip1pcThe dispersion in concentration and magnetization at
criticality between the different realizations is shown at a glance in
scattered plots as in Ref. \cite{Marques}. Each point in Fig. 4 represents a
single realization with magnetization at criticality ($M$) and concentration ($c 
$). Note that in both cases (TDIS and RDIS), the dispersion on the magnetization and on the
concentration decreases with L, but this is more clearly shown in the
thermal case. Fig. 4 shows clearly the difference in behavior between the 
{\bf random} and the {\bf thermal} cases, at least up to the values of $L$ considered.
\begin{figure}
\epsfxsize=\columnwidth\epsfysize=9cm\epsfbox{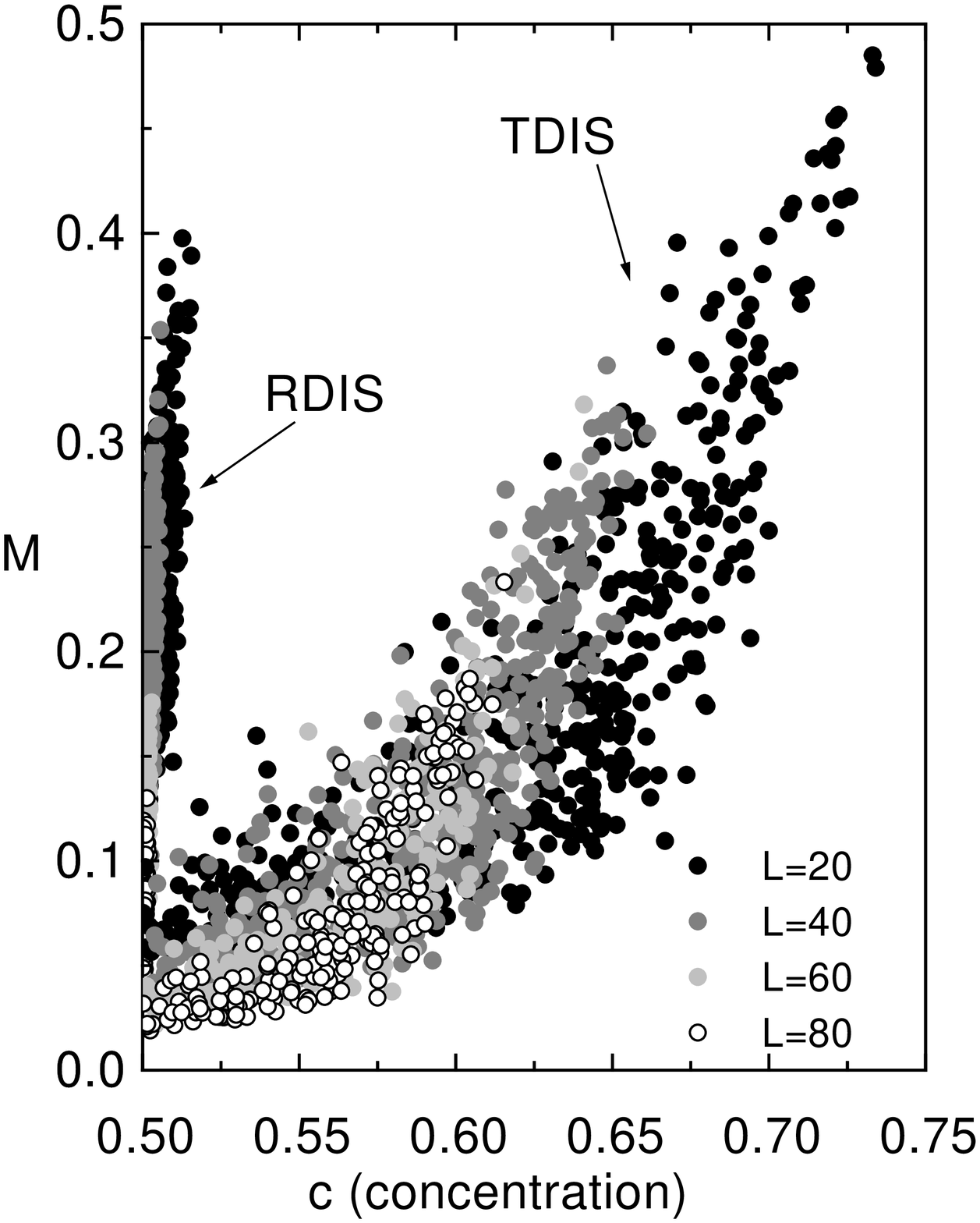}
\caption{Scattered plot of magnetization $M$ at the critical temperature vs.
concentration $c$ for the realizations considered of the random ($p=0.5$)
and the thermal dilutions. ($L=20,40,60,80$).}
\label{fig4}
\end{figure}

From Fig. 4 we can extract averaged values for the magnetization and the
inverse susceptibility, $\overline{M}$ and $\overline{\chi ^{-1}}$ (with $\chi
=\left\langle M^{2}\right\rangle $ at the critical point) for the TDIS. Both
averaged values are expected to fit the following scaling laws at
criticality: 
\begin{equation}
\overline{M(L)}\propto L^{-\beta /\nu }
\end{equation}
\begin{equation}
\overline{\chi ^{-1}(L)}\propto L^{-\gamma /\nu }
\end{equation}

Considering $(1/\nu )_{thermal}=a/2=0.985$, we can obtain from our data the
values of $\beta _{thermal}$ and $\gamma _{thermal}$ (see f.i. the fitting for
the inverse of the susceptibility in Fig. 5). We get $%
\beta _{thermal}=0.56\pm 0.05$ and $\gamma _{thermal}=1.91\pm 0.06$, very
close to the predicted values by Weinrib and Halperin \cite{Weinrib}. Using
the scaling relation: 
\begin{equation}
\alpha =2-2\beta -\gamma
\end{equation}
we obtain the following specific heat critical exponent: $%
\alpha _{thermal}=-1\pm 0.1$. Weinrib and Halperin give for LRC systems 
\cite{Weinrib} $\alpha =-1$, in good agreement with our result. An
analogous analysis has been performed but using the dispersion in
magnetization and inverse susceptibility instead of the averaged values. The
results are similar. The same study has been made for the random case (also
shown in Fig.5).
\begin{figure}
\epsfxsize=\columnwidth\epsfbox{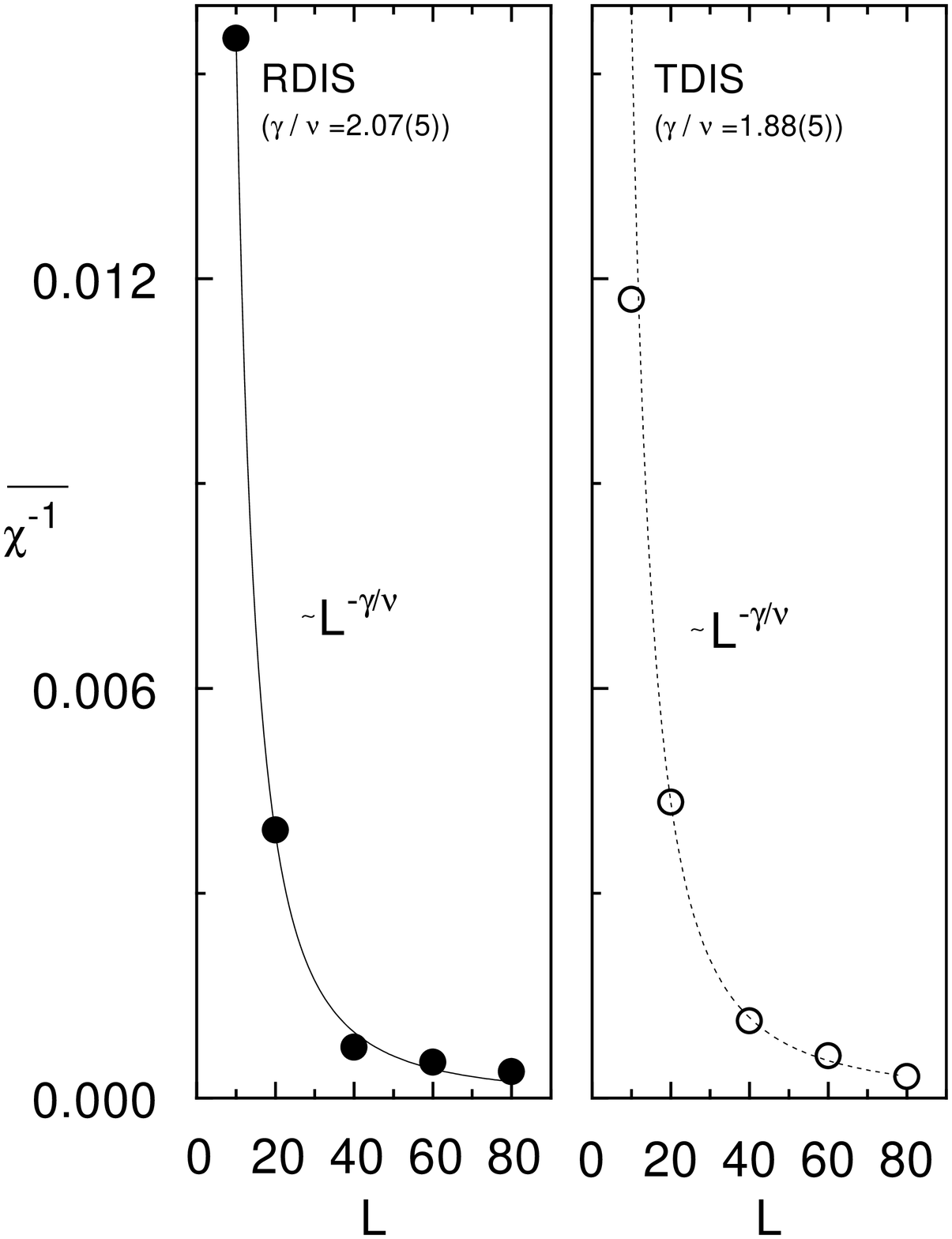}
\caption{Average value of the inverse of the susceptibility $\overline{%
\protect\chi ^{-1}(L)}$ vs. the lenght of the system $L$ for the random
dilution ($p=0.5$) case (black) and the thermal dilution case (white).
Continuous line and dashed line indicate fits to the random and the thermal
cases, respectively.}
\label{fig5}
\end{figure}

The critical exponents obtained are in agreement with those
of Ballesteros et al. \cite{BallesterosII} within our error bars.

The average concentration for the thermal system is also expected to show a
scaling law behavior given by: 
\begin{equation}
(\overline{c(L)}-0.5)\propto L^{-(\beta /\nu )_{3D}}
\end{equation}
where $(\beta /\nu )_{3D}\simeq 0.52$ gives the values corresponding to the
pure three dimensional Ising case, because in critically thermally diluted Ising systems,
vacancies are distributed with the same {\bf long range correlation} spin
distribution function as in the {\bf pure} case\cite{Marques}. The fit to the average
value of the concentrations shown in Fig. 4 give, for the thermal case, a
value $(\beta /\nu )_{3D}\simeq 0.55\pm 0.08$. This implies also a clear
difference between RDIS{\bf \ }and TDIS, since Ising systems with vacancies
randomly distributed are not expected to follow an scaling behavior with
$(\beta /\nu )_{3D}$. [Fitting to an scaling law the
results for RDIS gives an exponent around $1.4$, which implies a much faster
convergence to $c=0.5$].

\vskip1pc{\bf IV. Effective exponents and possible crossover for finite
systems} \vskip1pcThe difference between the universality class of RDIS and
TDIS can be detected also by means of the effective critical exponents. In
the case of the magnetization the effective critical exponent is defined by: 
\begin{equation}
\beta _{eff}\equiv \partial \log (M)/\partial \log (t)\quad (t=T_{c}^{i}-T)
\end{equation}
with $T_{c}^{i}$ the critical temperature of the {\bf particular realization} (i)
characterized by a maximum of the susceptibility ($\chi =\left\langle
M^{2}\right\rangle -\left\langle M\right\rangle ^{2}$). For $L\rightarrow
\infty $ and $t\rightarrow 0$, $\beta _{eff}=\beta $. Finite size effects
force the effective critical exponent to drop to zero before the critical
value is attained. However, since $\beta _{random}=0.3546$ (Ref. \cite
{BallesterosII} ), and $\beta _{thermal}$ is expected to be around $0.5$ (Ref. \cite{Weinrib} )
, effective critical exponents (for the thermal case)
may rise to values greater than 0.3546 and lower than 0.5, before finite
size effects appear, indicating the difference of universality class between
both kinds of systems. In the RDIS the effective critical exponent is
expected to be always lower than 0.3546. Monte Carlo simulations of
magnetization vs. temperature have been performed for randomly $(p=0.5)$ and
thermally diluted systems, with $L=80$. In Fig. 6 we show the results for $%
\beta _{eff}$ vs. $\log (t)$ for two samples of the TDIS and the RDIS type
respectively. The {\bf random} effective critical exponent is always under 0.35
and it seems to tend towards this value for large enough $L$, as expected,
but for {\bf thermal} systems the behavior is completely different: In the figure
the value of the critical thermal effective exponent is between 0.35 and
0.5. An analogous investigation has been done for different values of L and
different realizations. The same effect has been found for lower L values.
 However, we may note that in TDIS the effective critical exponent
arrives at the maximum in a very different way depending on the realization.
The reason is twofold: (1) the different {\bf disposition} of the vacancies in
each particular realization, and (2) the large differences in 
{\bf concentration} for the thermal case (the rise of $\beta _{eff}$ towards the
expected diluted universality value should be faster for c closer to 0.5). 
It may be noted that as the size $(L)$ of the sample increases the possibility of local
inhomogeneities in the TDIS realizations increases, giving rise to such
phenomena as pseudo double peaks in the susceptibility, reduced values for the
overall critical exponents, etc. However, it is important to remark than in
all realizations investigated the $\beta _{eff}$ values of the TDIS, (before
finite size effects take over), have been clearly superior to the $\beta $ for
the RDIS.

\begin{figure}
\epsfxsize=\columnwidth\epsfbox{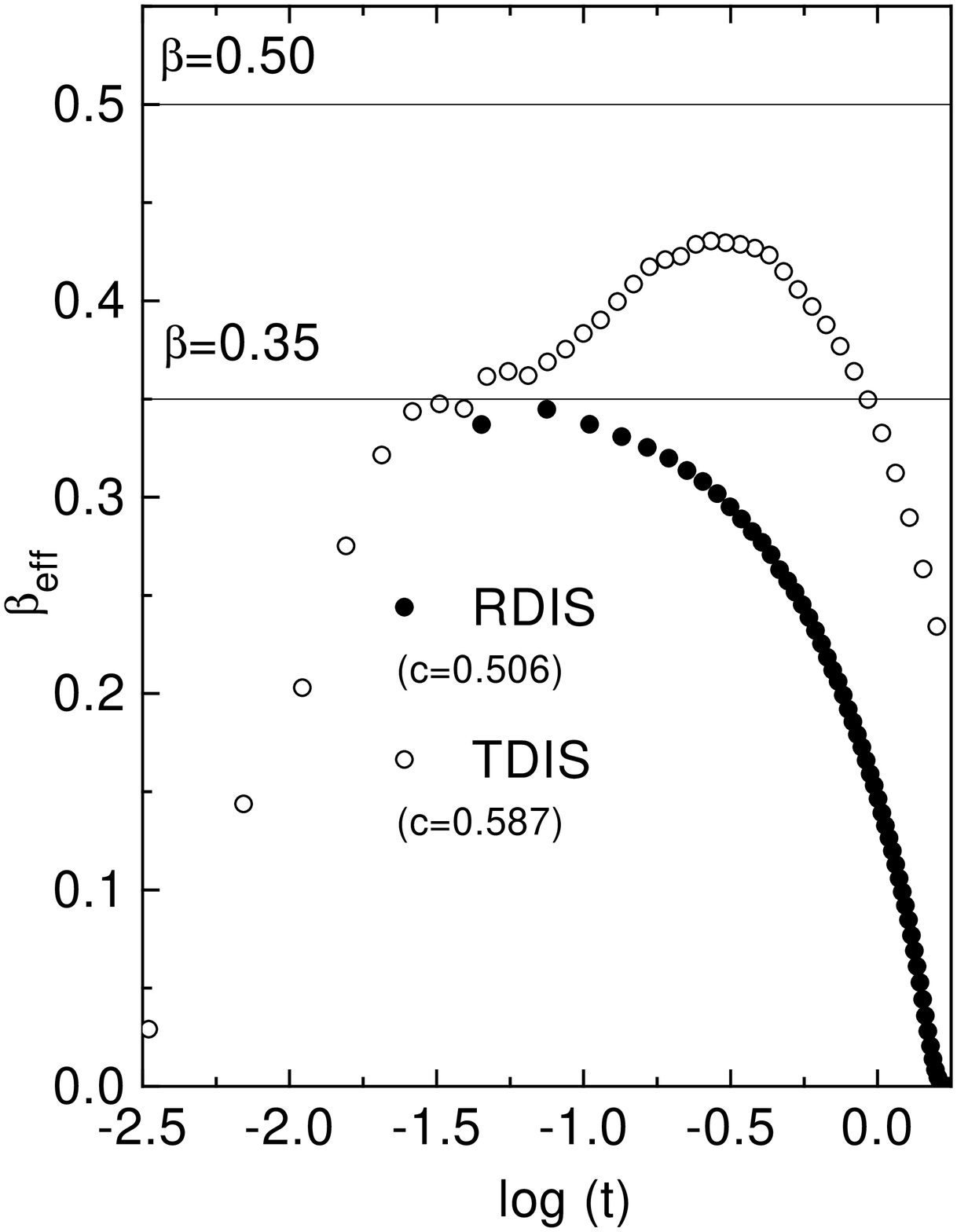}
\caption{Effective critical magnetization exponent $\protect\beta _{eff}$
vs. $log(t)$, with $t=T_{c}^{i}-T$, for the thermal dilution case (white)
and the random dilution ($p=0.5$) case (black). The realizations considered
correspond to $L=80$.}
\label{fig6}
\end{figure}

In Fig. 6 the effective critical exponent for the thermal case seems to
produce a crossover towards $\beta =0.35$ as the temperature approaches the
critical point, before finite size effects finally appear. In principle this
might be influenced by the indetermination in the measurement of the
particular critical temperature of the realization. However, this crossover
may point out that different length scales might be important at
different distance from the critical point. In principle characteristic
lengths such as the size of the fixed vacancies and spin clusters, the size
of the system itself and the thermal spin fluctuations might all play a role. The
possibility that for $L\longrightarrow \infty $ and $c=0.5$ the apparent
crossover could disappear all together should not be excluded, entailing no crossover
from critical thermal to random universality class.
\vskip1pc{\bf V. Concluding remarks} \vskip1pcTo summarize,
a new way to produce diluted Ising systems with a long range
correlated distribution of vacancies has been analyzed by means of Monte
Carlo calculations and finite size scaling techniques. We find an
universality class different from that reported for diluted Ising systems
with short range correlated disorder. Our systems may be included in the
universality class predicted by Weinrib and Halperin for $a\approx 2$. This
kind of thermal disorder had been already applied in percolation problems,
but as far as we know it had never been applied to magnetic systems, in
which long range correlated disorder has been previously produced mostly by random
distribution of lines or planes of vacancies. The present dilution procedure may be applicable
to systems where the long range correlated disorder is not due to
dislocations, preferential lines or planes of vacancies, but to systems
where the vacancies (points) are critically distributed in clusters as in the case of
order-disorder systems.

We thank H.E. Stanley for encourament and we are grateful to H.G.
Ballesteros, L.A. Fern\'{a}ndez, V. Mart\'{\i}n-Mayor, and A. Mu\~{n}oz
Sudupe for helpful correspondence. We thank P.A. Serena 
for helpful comments and for generous access to his computing facilities. We acknowledge
financial support from DGCyT through grant PB96- 0037 and from the Basque
Regional Government (J.I.).

\end{multicols}
\end{document}